\documentclass[%
reprint,amsmath,amssymb,aps,
]{revtex4-1}
\usepackage[colorlinks=true,urlcolor=blue,linkcolor=blue,citecolor=blue]{hyperref}
\usepackage{amsmath} 
\usepackage{amssymb}
\usepackage{epsf}
\usepackage{color}
\usepackage{dcolumn}
\usepackage{bm}

\usepackage{graphicx}
\usepackage[bottom]{footmisc}
\graphicspath{ {./images/} }

\begin{document}

\title{Dynamic effects of external axion fields in a system of many particles with spin}

\author{Mariya Iv. Trukhanova$^{1,2}$}
\email{trukhanova@my.msu.ru}

\author{Yuri N. Obukhov$^{2,3}$}
\email{obukhov@ibrae.ac.ru}

\affiliation{$^{1}$Theoretical Physics Department, Faculty of Physics,
Lomonosov Moscow State University, Leninskie Gory-1, 119991 Moscow, Russia \\
$^{2}$Theoretical Physics Laboratory, Nuclear Safety Institute, Russian Academy of Sciences,
B. Tulskaya 52, 115191 Moscow, Russia  \\
$^{3}$L. D. Landau Institute for Theoretical Physics, 142432 Chernogolovka, Russia}

\begin{abstract}
We develop the theoretical model that describes dynamic non-equilibrium effects of external inertial and axion fields in a system of particles with spin. The possibility of using the spin density and the current density of non-relativistic quantum particle systems for the detection of the hypothetical axion-like dark matter is discussed. The resulting closed system of dynamic equations encompasses the continuity equation, the momentum balance equation, and the spin density evolution equation, accounting for the influence of the spin-rotation coupling and the external axion fields. The new formalism opens up new perspectives for an experimental search of dark matter axions. 
\end{abstract}
\maketitle

\section{Introduction}

The strong CP problem, or the conservation of combined charge symmetry and parity symmetry in quantum chromodynamics, is one of the most important issues in theoretical physics \cite{PhysRevLett.38.1440, PhysRevLett.40.279}. A solution to the strong CP problem leads to the emergence of a new physical field and to the prediction of a new light pseudo-scalar particle, the axion, the mass of which is given by \cite{CHENG19881}
\begin{equation}
m_a\approx 6\cdot10^{-6}\,{\rm eV}\biggl(\,{\frac{10^{12}{\rm GeV}}{f_a}}\biggr),
\end{equation}
where $f_a$ is the decay constant of the axion field and has the dimension of energy. Besides, numerous other axion-like particles (ALPs) are considered as dark matter candidates, which are not associated with the strong CP problem, for example, in the framework of the string theory \cite{MARSH20161}.

One direction of searching for axions is based on the coherent mixing of axions and photons in a strong magnetic field, which is described by the Lagrangian density
\begin{equation}
L_{a\gamma\gamma} = -\,{\frac{g_{\gamma}\alpha}{\pi cf_a}}a(x)\bm{E}(x)\cdot\bm{B}(x),
\end{equation}
where $\alpha$ is the fine structure constant, $g_{\gamma}$ is a dimensionless model-dependent parameter of order unity \cite{SHIFMAN1980493}, and $a(x)$ is the axion field. Sikivie demonstrated \cite{PhysRevD.32.2988} that axions floating around in the halo of our galaxy or emitted by the Sun could resonantly convert into a monochromatic microwave signal in strong external electric and/or magnetic fields, with the photons produced to be captured in a high-quality factor tunable microwave cavity. To increase the signal, one of the cavity modes must be equal to the frequency of the axion signal \cite{PhysRevLett.51.1415}. The history of the microwave cavity experiment is quite extensive, with the first attempts going back to experiments at the Brookhaven National Laboratory \cite{PhysRevLett.59.839} and the University of Florida \cite{PhysRevD.42.1297}. The experiment with Rydberg-atom-cavity detectors (CARRACK) in Kyoto was aimed at reducing the system noise temperature \cite{TADA2006488}. The microwave cavity searches for dark-matter using the Kim-Shifman-Vainshtein-Zakharov (KSVZ) axion models were considered in Refs. \cite{PhysRevD.64.092003, PhysRevLett.104.041301}. The ADMX (Axion Dark Matter Experiment) and ADMX-HF (high resolution) Collaborations data acquisition channel is sensitive to such signals and was dedicated to the search for axion dark matter that fills our galactic halo and has covered the axion mass range of $1.9\,\mu$eV$/c^2 < m_a < 3.65 \,\mu$eV$/c^2$\,, \cite{PhysRevD.84.121302}. 

The search for solar axions is realized with helioscopes based on powerful magnets, which are oriented to the Sun and coupled to X-ray detectors. The conversion of axions to photons in a magnetic field can be reinforced with wire arrays \cite{PhysRevLett.98.172002, PhysRevD.50.4744} or dielectric plates to modify the mode structure of the electromagnetic field \cite{Dieter2013}.  Since 2003, the CERN experiment CAST (CERN Axion Solar Telescope) has been realized using the magnet with a magnetic field of up to 9T. An upper limit on the axion-photon coupling was obtained of $g_{a\gamma\gamma} < 8.8\cdot 10^{-11}$GeV$^{-1}$, \cite{Andriamonje2007}. The more sensitive experiment IAXO, which is based on a more powerful magnet, has a signal-to-noise ratio that is five orders of magnitude better than that of CAST \cite{Lakic2020}. This made it possible to measure the constant of the axion-photon coupling $\sim 5\cdot 10^{-12}$GeV$^{-1}$. Of great interest is working with laboratory sources of axions and ALPs. The light shining through walls (LSW) experiments, based on the direct and inverse Primakoff effect ($\gamma\longrightarrow a\longrightarrow\gamma$), have been carried out in several scientific groups \cite{PhysRevLett.101.120401, PhysRevLett.100.080402, EHRET2010149, PhysRevLett.99.190403, PhysRevD.80.072004, JAECKEL2008509, RBahre2013}.

Another promising direction of the experimental search for dark matter by making use of the nuclear magnetic resonance techniques in the Cosmic Axion Spin Precession Experiment (CASPEr) \cite{PhysRevD.88.035023,PhysRevD.89.043522,PhysRevD.92.105025,PhysRevD.103.116021} is based on the direct coupling of axion-like particles to the spin of fermions. The search for the influence of the axion wind due to the interaction of axions with the nucleon spin, that should occur around the axion field gradient, is investigated in the CASPEr-Wind experiment over a wide range of axion masses $10^{-9}$eV$/c^2$--$10^{-14}$eV$/c^2$, \cite{PhysRevX.4.021030,Garcon_2018}. The interaction of fermion's spin with a cold axion field (``axion wind'') is similar to the interaction with a pseudomagnetic field $\sim \bm{\nabla}a$ that produces the spin torque ``axion wind'' \cite{Berlin}. Manifestations of this pseudomagnetic field can be realized in magnetically ordered media \cite{Gramolin, PhysRevLett.124.171801, PhysRevLett.113.161801}. For example, a ferromagnetic axion haloscope, based on the interaction of axions with electron spins, can be used to search for dark matter \cite{PhysRevLett.124.171801}. The experimental scheme with toroidal magnets with ferromagnetic powder cores made of an iron–nickel alloy \cite{Gramolin} is based on a modification of Maxwell's equations in the presence of axions that mix with a static magnetic field to produce an oscillating magnetic field. The action of a pseudomagnetic field generated by axions can also manifest itself as a result of the excitation of magnons \cite{BARBIERI1989357, k6pg-bkwh} or in experiments on the resonant rotation of nuclear spins \cite{PhysRevD.97.055006}. It is also important to take into account Earth's rotation. The extension of recent studies from the flat Minkowski geometry to curved spacetime manifolds \cite{ObukhovYu} contributes to the discussion of the possible new role of a precessing spin as an ``axion antenna'', revealing the influence of the gravitational/inertial and axion fields in the ultra-sensitive high-energy experiments with polarized particles. It is expected that the study of the spin dynamics in accelerators and storage rings will be able to enhance the pseudomagnetic field effects at ultra-relativistic velocities \cite{SilenkoAJ,Nikolaev,UFN:2023}. 

Another promising area of research is related to the study of long-range forces mediated by axions, which may lead to deviations from the gravitational law $1/r^2$ \cite{PhysRevD.30.130}. In this case, an interaction energy between the two fermions can be represented by the sum of `monopole-monopole', `monopole-dipole' \cite{PhysRevLett.98.081101, PhysRevLett.120.161801} and `dipole-dipole' \cite{PhysRevA.95.032505, PhysRevLett.115.201801} interactions. The existence of a new stable massless pseudoscalar boson (arion), corresponding to spontaneous breaking of the chiral leptonic symmetry, was put forward in Refs. \cite{ANSELM198239, ANSELM1982161}. A massless axion can exist even in the absence of a massive axion. The long range forces mediated by arions are equivalent to the magnetic interactions between spins, however without an electric current. Therefore, superconducting screens should not shield this field, unlike the magnetic one.     

Despite numerous works on axion-fermion coupling \cite{PhysRevLett.133.191801, dxvs-stdk}, the non-relativistic collective dynamics of quantum particles with spin and the evolution of the spin current in the pseudoscalar field of dark matter axions are still poorly understood. The goal of our paper is to construct a theoretical model of collective spin dynamics of quantum systems, taking into consideration the axion-fermion coupling.

This paper is organized as follows. In Sec.~\ref{Preliminaries}, the Dirac equation is recast into the  Pauli-Schr\"odinger form by the Foldy-Wouthuysen transformation for the Dirac Hamiltonian, the structure of the Hamiltonian includes the spin-axion and rotation-axion coupling. In Sec.~\ref{MPfSAC} the quantum hydrodynamic model for the spin and current dynamics of a system of particles in the pseudoscalar field of dark matter axions is derived, including the basic definitions and major derivation steps. In Sec.~\ref{Discussion} a brief summary and discussion of the obtained results is presented.

\section{Preliminaries}\label{Preliminaries}

In general relativity, both the gravitational and inertial effects are encoded in the metric and connection structures on a four-dimensional curved space-time, where local holonomic coordinates $x^i=(t,x^a)$ are introduced \cite{PhysRevD2009}. The metric defines the distances and lengths of the vectors and the angles between the vectors, while the connection determines the parallel transport and introduces covariant derivatives. The components of an arbitrary spacetime metric $g_{ij}$ are most conveniently described in terms of Arnowitt-Deser-Misner parametrization \cite{PhysRevD2011}: 
\begin{align}
g^{00} &= \frac{1}{c^2V^2},\quad g^{0a} = \frac{K^a}{cV^2},\quad g^{ab}=-\underline{g}^{ab}+\frac{K^aK^b}{V^2},\\
g_{00} &= c^2(V^2-K^2), \quad g_{0a} = c\underline{g}_{ab}K^b,\quad
g_{ab} = -\underline{g}_{ab},
\end{align}
where $\underline{g}_{ab} = \delta_{\hat{c}\hat{d}}W^{\hat{c}}{}_aW^{\hat{d}}{}_b$ and $K^2 = \underline{g}_{ab}K^aK^b$.

Let us consider the coupling of the axion field to a Dirac fermion. The dynamics of a fermion field $\Psi$ with spin ${\frac 12}$, rest mass $m$ and electric charge $q$, which is coupled with an external dynamical pseudo-scalar axion field $a(x)$, is derived from the Lagrangian 
\begin{align}
{\cal L}_{af} =& \,\,\frac{i\hbar}{2}\left(\bar{\Psi}\gamma^{\mu}D_{\mu}\Psi
- D_{\mu}\bar{\Psi}\gamma^{\mu}\Psi\right)-mc\bar{\Psi}\Psi\nonumber\\
& -\,{\frac{\hbar}{2}}{\frac{g_f}{f_{a}}}\Bar{\Psi}\gamma^{\alpha}\gamma_5\Psi(e^i_{\alpha}\partial_ia),\label{LD}
\end{align}
where $\hbar$ is a Planck constant, $D_{\mu}$ is a spinor covariant derivative, $f_a$ is a quantity with the dimension of $a$, called the axion decay coupling constant, and the model dependent dimensionless constant $g_f\sim 1$ \cite{RevModPhys.93.015004}. In order to discuss the Dirac spinors, we need orthonormal frames. The  Dirac $4\times 4$ matrices have their usual form, namely \cite{BS},
\begin{align}
    \gamma^{\hat{0}} &= \frac{1}{c}\begin{pmatrix}
        1 & 0 \\ 0 & -1
    \end{pmatrix}, \qquad \alpha^{\hat{a}}=\begin{pmatrix}
        0 & \sigma^a \\ \sigma^a & 0
    \end{pmatrix}, \\
    \gamma_5 &= \begin{pmatrix}
        0 & -1 \\ -1 & 0
    \end{pmatrix}, \qquad \Sigma^{\hat{a}}=\begin{pmatrix}
        \sigma^a & 0 \\ 0 & \sigma^a
    \end{pmatrix}.
\end{align}
The minimal coupling of a fermion wave function to the electromagnetic field $A_i$ and the gravitational field $(e_i^{\mu}, \Gamma_i{}^{\beta\gamma})$ potentials \cite{ObukhovYu} is encoded in the spinor covariant derivative
\begin{equation}
D_{\mu} = e^i_{\mu}\left(\partial_i - {\frac{iq}{\hbar}}A_i
+ \frac{i}{4}\Gamma_i{}^{\beta\gamma}\sigma_{\beta\gamma}\right),
\end{equation}
where $\Gamma_i{}^{\beta\gamma}$  are the Lorentz connection coefficients and $\sigma^{\mu\nu}=\frac{i}{2}\left(\gamma^{\mu}\gamma^{\nu} - \gamma^{\nu}\gamma^{\mu}\right)$ are the Lorentz group generators in the anholonomic coordinates. The Dirac conjugate spinor is defined by $\Bar{\Psi} = \Psi^{\dagger}\beta$, where $\beta=c\gamma^{\hat{0}}$, and its covariant derivative reads as
\begin{equation}
D_{\alpha}\Bar{\Psi}=e^i_{\alpha}\left(\partial_i\Bar{\Psi}+\frac{iq}{\hbar}A_i\Bar{\Psi}
-\frac{i}{4}\Gamma_i{}^{\beta\gamma}\Bar{\Psi}\sigma_{\beta\gamma}\right).
\end{equation}
We choose the Schwinger gauge \cite{PhysRevD2011,PhysRevD2009} for the tetrad components:
\begin{align}
e^{\hat{0}}_{i} &= V\delta^0_i, \qquad e^{\hat{a}}_i = W^{\hat{a}}{}_{b}\left(\delta^b_i-cK^b\delta^0_i\right),\\
e^i_{\hat{0}} &= \frac{1}{V}\left(\delta^i_0 + \delta^i_bcK^b\right),\qquad e^i_{\hat{a}}=\delta^i_bW^b{}_{\hat{a}}, 
\end{align}
In order to guarantee the hermiticity of the Hamiltonian, one needs to rescale the spinor wave function:
\begin{equation}\label{rescale}
    \psi = \left(\sqrt{-g}e^0_{\hat{0}}\right)^{1/2}\Psi.
\end{equation}
As a result, the Dirac equation, derived from the action functional with the Lagrangian (\ref{LD}) for the rescaled wave function \eqref{rescale}, is recast into the Schr\"odinger form $i\hbar\partial_t\psi=\hat{H}\psi$ with the Hermitian Hamiltonian  
\begin{align}
\hat{H} =&\,\, \beta Vmc^2+q\phi + {\frac{c}{2}}\left(\pi_bVW^b{}_{\hat{a}}\alpha^{\hat{a}}
+ \alpha^{\hat{a}}VW^b{}_{\hat{a}}\pi_b\right)\nonumber\\
& + \frac{c}{2}\left(\bm{K}\cdot\bm{\pi} + \bm{\pi}\cdot\bm{K}\right)
+ {\frac{\hbar c}{4}}\left(\bm{\Xi}\cdot\bm{\Sigma} - \Upsilon\gamma_5\right)\nonumber\\
&+ {\frac{\hbar g_f}{2f_a}}\left(\gamma_5\left(\dot{a} + c\bm{K}\cdot\bm{\nabla}a\right)
+ cVW^b{}_{\hat{a}}\alpha^{\hat{a}}\gamma_5\partial_ba\right).\label{hamiltonian0}
\end{align}
Here the kinetic momentum operator is $\bm{\pi} = \{\pi_a\} = - i\hbar\partial_a - q A_a$; $A_i=(-\phi, A_a)$ are scalar and vector potentials of an external electromagnetic field, and
\begin{equation}
\Xi^a = {\frac Vc}\,\epsilon^{abc}\Gamma_{\hat{0}\hat{b}\hat{c}},\qquad
\Upsilon = V\,\epsilon^{abc}\Gamma_{\hat{a}\hat{b}\hat{c}}.
\end{equation}
The Dirac equation with the Hamiltonian (\ref{hamiltonian0}) acts on a four-component spinor wave function.

\subsection{Dirac fermion in rotating frame}\label{rotframe}

We now specialize to the flat spacetime in the rotating frame:
\begin{equation}
V = 1,\qquad W^{\hat{a}}{}_b = \delta^a_b,\qquad \bm{K} = -\,{\frac{1}{c}}\,\bm{\omega}\times\bm{r},
\end{equation}
where $\bm{\omega}$ is the angular velocity. In this case, the Hamiltonian (\ref{hamiltonian0}) reduces to 
\begin{align}
\hat{H} =&\,\, \beta mc^2+q\phi + c\bm{\alpha}\cdot\bm{\pi}
-\bm{\omega}\cdot(\bm{r}\times\bm{\pi}) - \frac{\hbar}{2}\bm{\omega}\cdot\bm{\Sigma}\nonumber\\
&+ \gamma_5\left(\lambda - {\frac{1}{c}}\,\bm{\lambda}\cdot(\bm{\omega}\times\bm{r})\right)
- \bm{\lambda}\cdot\bm{\Sigma},\label{hamiltonian}
\end{align}
where we introduced the notation for the derivatives of the axion field
\begin{equation}\label{lam}
\bm{\lambda}=\frac{c\hbar g_f \bm{\nabla}a}{2f_a}, \qquad \lambda=\frac{\hbar g_f\Dot{a}}{2f_a}.
\end{equation}

\subsection{Foldy-Wouthuysen transformation}\label{FWT}

Let us derive the Pauli-Schr\"odinger equation in a rotating frame by the Foldy-Wouthuysen (FW) transformation, which is described in detail in Ref. \cite{PhysRevD2011, PhysRevA.77.012116}. These transformations are unitary and block off the diagonalization of the Dirac Hamiltonian. In the first step, we define the even ${\cal E}$ and odd ${\cal O}$ parts of the Hamiltonian in the form
\begin{equation}
    H=\beta mc^2+{\cal E}+{\cal O},
\end{equation}where
\begin{align}
{\cal E} &= q\phi-\frac{\hbar}{2}\bm{\omega}\cdot\bm{\Sigma}
- \bm{\lambda}\cdot\bm{\Sigma} - \bm{\omega}\cdot(\bm{r}\times\bm{\pi}),\\
{\cal O} &= c\bm{\alpha}\cdot\bm{\pi} + \gamma_5
\left(\lambda - {\frac{1}{c}}\,\bm{\lambda}\cdot(\bm{\omega}\times\bm{r})\right).
\end{align}
The diagonalization of the Hamiltonian is based on a low-energy expansion of the Hamiltonian $H'=UHU^{\dagger}-U\partial_tU^{\dagger}$ into an exponential series of $1/m$, which leads to the vanishing of the odd parts of $H'$, where $U = \exp(-\frac{i\beta{\cal O}}{2mc^2})$. The expansion of Hamiltonian $H'$ up to the order of $1/m^2$ leads to the expression
\begin{equation}\label{H2}
H'=\beta\left(mc^2+\frac{{\cal O}^2}{2mc^2}\right) + {\cal E}
-\frac{1}{8m^2c^4}\left[{\cal O},[{\cal O},{\cal E}]+i\hbar\Dot{{\cal O}}\right].  
\end{equation}
After a set of simple calculations, the term that is proportional to $1/mc^2$ in Eq. \eqref{H2} becomes
\begin{equation}\label{beta Omega}
\frac{\beta{\cal O}^2}{2mc^2}=\beta\left(\frac{\bm{D}^2}{2m}-\mu\bm{\Sigma}\cdot\bm{B}
-\frac{\tilde{\lambda}^2}{mc^2}\right),
\end{equation}
Here $\mu = {\frac {\hbar q}{2mc}}$ is the magnetic moment of the fermion, and
\begin{align}
\bm{D} &= \bm{\pi} - {\frac{1}{c}}\,\tilde{\lambda}\,\bm{\Sigma},\\
\tilde{\lambda} &= \lambda - {\frac{1}{c}}\,\bm{\lambda}\cdot(\bm{\omega}\times\bm{r}).
\end{align}
The contribution of the spin-axion and rotation-axion coupling effects in the second term $-\frac{1}{8m^2c^4}[{\cal O},[{\cal O},{\cal E}]+i\hbar\Dot{{\cal O}}]$ can be derived in the form
\begin{align}
&-\frac{1}{8m^2c^4}\biggl[{\cal O},[{\cal O},{\cal E}]+i\hbar\Dot{{\cal O}}\biggr]=-\frac{\hbar}{4m^2c^2}(\bm{\nabla}\times\bm{\lambda})\cdot\bm{\pi}\nonumber\\
&-\frac{1}{4m^2c^2}\Sigma^a\lambda^b\{\pi^b,\pi^a\}
+\frac{1}{4m^2c^2}(\bm{\Sigma}\cdot\bm{\lambda})\{\bm{\pi},\bm{\pi}\}\nonumber\\
&-\frac{\hbar^2}{8m^2c^2}\bm{\Sigma}\cdot\nabla^2\bm{\lambda}-\frac{\hbar q}{2m^2c^3}\bm{\lambda}\cdot\bm{B}
+\frac{\hbar^2}{8m^2c^3}(\bm{\Sigma}\cdot\bm{\nabla})\dot{\tilde{\lambda}}\nonumber\\
&-\frac{\hbar^2}{8m^2c^3}\bm{\alpha}\cdot(\bm{\omega}\times\bm{\nabla}\tilde{\lambda})
+\frac{\hbar^2}{8m^2c^3}\alpha^b\bm{\omega}\cdot(\bm{r}\times\partial_b\bm{\nabla}\tilde{\lambda}).\label{Ha}
\end{align}
In the expression above and in the further derivations we leave out terms which are proportional to the square of the coupling constant with fermions $g_f$. Following this approach, the third term in \eqref{beta Omega} $\sim \tilde{\lambda}^2/c^2$ will also be discarded in subsequent calculations. 

In the next step, we obtain the Pauli-Schr\"odinger equation for the upper component of the Dirac spinor in the low-energy limit. Neglecting the rest energy term in Eq. \eqref{H2}, the Pauli-Schrodinger equation for the upper component of the Dirac spinor, namely, 2-component wave function, in the rotating frame of reference, in magnetic and axion fields, has the form $i\hbar\frac{\partial\psi}{\partial t}=\hat{H}\psi,$
\begin{equation}\label{total H}
\hat{H} = \hat{H}_0 + \hat{H}_{\rm zeeman} + \hat{H}_{\rm rot} + \hat{H}_{\rm SO} + \hat{H}_{\rm darwin} + \hat{H}_{\rm ax}.
\end{equation}
The spin-independent part of Hamiltonian $\hat{H}_0$ contains the contribution of the kinetic energy, the potential energy of the charge in an external electric field with a scalar potential $\phi$, and the potential energy in the axion field 
\begin{equation}
\hat{H}_0 = {\frac 1{2m}}\left(\hat{\bm{\pi}} - \frac{\tilde{\lambda}}{c}\hat{\bm{\sigma}}\right)^2
+ q\phi - \frac{\tilde{\lambda}^2}{mc^2}.
\end{equation}
The Zeeman energy of the magnetic moment in an external magnetic field, axion-magnetic field coupling, and spin-axion coupling effects are included in the second term of the Hamiltonian \eqref{total H}
\begin{equation}
\hat{H}_{\rm zeeman} = -\,\mu\bm{B}\cdot\left(\hat{\bm{\sigma}}+\frac{\bm{\lambda}}{mc^2}\right)
-\bm{\lambda}\cdot\hat{\bm{\sigma}}.
\end{equation}
The non-inertial effects of the coupling of angular momentum with rotation velocity $\bm{\omega}$ and spin-rotation interaction are characterized by
\begin{equation}
    \hat{H}_{\rm rot}=-\,\bm{\omega}\cdot\left(\hat{\bm{r}}\times\hat{\bm{\pi}}+\frac{\hbar}{2}\hat{\bm{\sigma}}\right).
\end{equation}
The spin-orbital coupling and Darwin term have the well-known form
\begin{align}
\hat{H}_{\rm SO} &= \frac{q\hbar}{8m^2c^2}\hat{\bm{\sigma}}\cdot
\left(\hat{\bm{\pi}}\times\bm{E}'-\bm{E}'\times\hat{\bm{\pi}}\right),\\
\hat{H}_{\rm darwin} &=\frac{q\hbar^2}{8m^2c^2}\hat{\bm{\nabla}}\cdot\bm{E}'.
\end{align}
Here $\bm{E}'=\bm{E}+(\bm{\omega}\times\bm{r})\times\bm{B}$ is the electric field in a rotating reference frame. The Hamiltonian of the axion field $\hat{H}_{\rm ax}$ has a complex form and is determined by Eq. \eqref{Ha}. Expansion to the order inverse to the square of the particle mass $1/m^2$ leads to the appearance of the spin-orbit interaction and the Darwin term. In a rotating reference frame, we find that the electric field $\bm{E}$ is modified by an additional term $(\bm{\omega}\times\bm{r})\times\bm{B}$, which is consistent with the theory.

\section{\label{MPfSAC} Many-particle model for spin-axion coupling}

In Sec.~\ref{FWT} we derived the Hamiltonian for spin-axion coupling in the rotating frame of reference. Let us generalize this model to the case of many-particle systems with spins. For this purpose, we use the method of many-particle quantum hydrodynamics (MPQH), which was developed for various physical systems, in particular, for quantum plasmas \cite{Andreev_Trukhanova_2018}, for scalar and spinor Bose-Einstein condensates \cite{Mosaki_2021} and even for the description of ferroelectricity of spin origin in magnetoelectric materials \cite{Trukhanova_2024, Andreev_2024}.

Here we apply this method to describe the axion-fermion interactions in a many-particle system. We are interested in the evolution of collective variables such as the concentration of particles, the spin density, and the current density vectors. In addition, we consider the behavior of neutral particles with spins in the rotating frame of reference in the absence of an external electric field and describe their non-relativistic dynamics.   

In the lowest order of expansion from \eqref{total H}, neglecting contributions proportional to the square of the axion-fermion coupling constant, we construct the many-particle non-relativistic Hamiltonian according to the MPQH method
\begin{align}
  \hat{H} = \sum_p^N\biggl(\frac{\hat{\bm{j}}_p^2}{2m_p}-\bm{\lambda}_p\cdot\hat{\bm{\sigma}}_p-\frac{\hbar}{2}\bm{\omega}_p\cdot\hat{\bm{\sigma}}_p-\mu_p\hat{\bm{\sigma}}_p\cdot\bm{B}_p\nonumber\\
-\frac{\tilde{\lambda}_p}{c}(\bm{\omega}_p\times\hat{\bm{r}}_p)\cdot\hat{\bm{\sigma}}_p
-\frac{m_p}{2}(\bm{\omega}\times\hat{\bm{r}}_p)^2\biggr),\label{FW}
\end{align}
where we have introduced the effective momentum operator in the form
\begin{equation}
\hat{\bm{j}}_p=\hat{\bm{\pi}}_p-\frac{1}{c}\biggl(\lambda_p - {\frac{1}{c}}\,\bm{\lambda}_p\cdot
(\bm{\omega}_p\times\hat{\bm{r}}_p)\biggr)\hat{\bm{\sigma}}_p-m_p(\bm{\omega}_p\times\hat{\bm{r}}_p),
\end{equation}
and $\hat{\bm{\pi}}_p = -i\hbar\hat{\bm{\nabla}}_p - q\bm{A}_p$.

Let us analyze the structure of the many-particle Hamiltonian \eqref{FW} in more detail. The first term represents the kinetic energy of particles. The third term is the rotation energy, which can be rewritten as $-\mu_p\bm{\sigma}_p\cdot\bm{B}_{\omega}$ in terms of the Barnett field $\bm{B}_{\omega} = {\frac {m_pc}{q_p}}\,\bm{\omega}$ as an artificial magnetic field due to mechanical rotation, thus providing the quantum mechanical origin of the Barnett effect. The second, fourth and fifth terms represent the axion-spin coupling energy, the Zeeman energy of magnetic moment in an external magnetic field, the energy of interaction of spin and mechanical angular momentum in the presence of an axion field, and the sixth term is the potential energy of rotation with the angular velocity $\bm{\omega}$ in the non-inertial frame, respectively.

\subsection{Spin evolution equation}

For a system of $N$ particles, the many-particle wave function $\psi_s(R,t) = \psi_{s_1,\dots,s_N}(\bm{r}_1,\dots,\bm{r}_N,t)$ depends on the $3N$ space variables $R = \{\bm{r}_1,\dots,\bm{r}_N\}$, the time $t$, and the spin variables $S = \{s_1,\dots,s_N\}$. In the method of many-particle quantum hydrodynamics, one constructs macroscopic physical variables $\bm{f}(\bm{r},t)$ as the quantum-mechanical average values of the corresponding operator $\hat{\bm{f}}$ of the physical quantity based on the many-particle wave function of the quantum state. We will use the general notation for the averaging
\begin{align}
\bm{f}(\bm{r},t) = \langle\psi^{\dagger}_s\hat{\bm{f}}\psi_s\rangle 
= \sum_{S}\int dR\,\psi^{\dagger}_s(R,t)\hat{\bm{f}}\psi_s(R,t),\label{notation}
\end{align}
where the integral is taken over the $3N$-dimensional configuration space with the volume element $dR = \prod_{i = 1}^N\,d\bm{r}_i$. The typical structure of the averaged operator density is $\hat{\bm{f}} = \sum_p^N\delta(\bm{r}-\bm{r}_p)\hat{\bm{f}}_p$, where $\hat{\bm{f}}_p$ acts on a $p$-th particle. The important particular examples include the particle number density $n = \langle\psi_s^{\dagger}\hat{n}\psi_s\rangle$ determined by the concentration operator $\hat{n}=\sum_p\delta(\bm{r}-\bm{r}_p)$, and the spin density vector that arises as the quantum average  $\bm{s} = \langle\psi^{\dagger}_s\hat{\bm{\sigma}}\psi_s\rangle$ of the spin density operator $\hat{\bm{\sigma}}=\sum_{p}\delta(\bm{r}-\bm{r}_p)\hat{\bm{\sigma}}_p$. 
Explicitly, the spin density vector can be defined as
\begin{equation}\label{spin density}
\bm{s}(\bm{r},t)=\sum_{S}\int dR \sum_p^N\delta(\bm{r}-\bm{r}_p)\psi^{\dagger}_s(R,t)\hat{\bm{\sigma}}_p\psi_s(R,t).
\end{equation}
The evolution of spin density \eqref{spin density} is determined by the evolution of the spinor wave function $\psi_s(R,t)$ that encodes information about a quantum system of particles. The calculation of the spin density evolution requires the Hamiltonian \eqref{FW} for a system of particles in an external axion field. 

Consider the time derivative of the microscopic definition \eqref{spin density} to determine the equation for the evolution of the spin density. The time derivatives of many-particle wave functions can be expressed in terms of the Hamiltonian according to the Pauli-Schr\"odinger equation $i\hbar\partial_t\psi_s=\hat{H}\psi_s$. Separating the contribution of the kinematic terms and the terms describing the interaction with the external fields, we derive the dynamical equation for the spin density $\bm{s}(\bm{r},t)=s^{a}(\bm{r},t)$
\begin{align}
\partial_t\bm{s} + \nabla_{b}j^{ab}_s &= \frac{2\mu}{\hbar}\bm{s}\times\bm{B}+\overbrace{\,\bm{s}(\bm{r},t)\times\bm{\omega}(t)}^{\text{spin-rotation coupling}}\nonumber\\
\overbrace{+\,\frac{g_fc}{f_a}\bm{s}\times\bm{\nabla}a}^{\text{spin-axion coupling}}&+\overbrace{\,\frac{g_f\Dot{a}}{cf_a}\bm{\omega}\cdot\Lambda^{bb}-\,\frac{g_f\Dot{a}}{cf_a}\omega^b\cdot\Lambda^{ab}}^{\text{spin-axion-rotation coupling}}\nonumber\\
\overbrace{+\,\frac{g_f\dot a}{cf_a}
  \epsilon^{abc}j^{bc}_s}^{\text{spin current-axion coupling}}&-\overbrace{\,\frac{g_f}{cf_a}\epsilon^{dmn}\nabla^da\cdot\omega^m\cdot\Omega^{an},}^{\substack{\text{spin current-axion} \\ \text{-rotation coupling}}}\label{spin equation}
\end{align}
where the microscopic representation of the spin current density is obtained during the derivation of equation \eqref{spin equation}
\begin{equation}
j^{ab}_s(\bm{r},t)=\Bigl\langle\frac{1}{2m_p}\left\{(\hat{\sigma}^{\ a}_p\hat{j}^{\ b}_p\psi_s)^{\dagger}
\psi_s + \psi^{\dagger}_s(\hat{\sigma}^{\ a}_p\hat{j}^{\ b}_p\psi_s)\right\}\Bigr\rangle,\label{spin current density}
\end{equation}
and the operator $\hat{\bm{j}}_p$ for the neutral spinning particles has the form
\begin{align}
\hat{\bm{j}}_p = - i\hbar{\hat{\bm{\nabla}}}_p\overbrace{-m_p\bm{\omega}_p\times\hat{\bm{r}}_p}^{\text{rotational velocity}}\quad\quad\quad\quad\quad\quad\quad\nonumber\\
-\overbrace{\frac{\hbar g_f}{2cf_a}\biggl(\dot{a}_p-\hat{\bm{\nabla}}a_p\cdot(\bm{\omega_p\times\hat{\bm{r}}}_p)\biggr)\hat{\bm{\sigma}}_p.}^{\substack{\text{spin-dependent velocity} \\ \text{due to the spin-axion coupling}}}
\end{align}
We also introduce a set of new quantities $\Lambda^{ab}(\bm{r},t)$, $\Omega^{an}(\bm{r},t)$ that appear in the non-inertial rotating frame
\begin{align}\label{Lam}
\Lambda^{ab} &= \Bigl\langle \psi_s^{\dagger}(R,t)\hat{\sigma}_p^{\ a}\hat{r}_p^{\ b}\psi(R,t)\Bigr\rangle,\\
\Omega^{an} &= \epsilon^{abc}\Bigl\langle \frac{1}{2m_p}\left\{(\hat{\sigma}^{\ b}_p
\hat{j}^{\ c}_p\psi_s)^{\dagger}\hat{r}_p^{\ n}\psi_s + h.c\right\}\Bigr\rangle.
\end{align}
Let us discuss the structure of the equation \eqref{spin equation}. The first term on the right hand side of the spin evolution equation \eqref{spin equation} is the torque acting on the magnetic moment in the magnetic field as a result of the ordinary Zeeman energy. The second and third terms represent torques resulting from spin-rotational coupling and spin-axion coupling in a gradient axion field $\bm{\nabla}a(\bm{r},t)$ or when the axion field is non-uniformly distributed in space. The fourth and fifth torques follow from the axion-spin-rotation coupling in the time-varying axion field $\Dot{a}(\bm{r},t)=\partial_ta(\bm{r},t)$. The sixth term represents the dynamical effect of the coupling of the time-varying axion field with the spin current density tensor $j^{ab}_s(\bm{r},t)$.

\subsection{Particle dynamics in the non-inertial frame}

In this section, we obtain an equation to describe the non-equilibrium dynamics of systems of neutral particles in non-inertial frame and external axion field. For this purpose, in the first step, we introduce the concentration of particles in the neighborhood of a space point $\bm{r}$. If we define the concentration of particles as the quantum average of the concentration operator in the coordinate representation $\hat{n}=\sum_p\delta(\bm{r}-\bm{r}_p)$, we can define 
\begin{align}
n(\bm{r},t) &= \sum_S\int dR\sum_p\delta(\bm{r}-\bm{r}_p)\psi_s^{\dagger}(R,t)\psi_s(R,t)\nonumber\\
&= \biggl\langle\psi_s^{\dagger}(R,t)\psi_s(R,t)\biggl\rangle.\label{concentration}
\end{align}
Considering the time derivative of the particle concentration, we find that the time derivative acts on the wave functions being integral in the microscopic definition of concentration \eqref{concentration}. Using the Pauli-Schr\"odinger equation with the Hamiltonian \eqref{FW}, we replace the time derivatives of the wave functions $\partial_t\psi_s^{\dagger}, \partial_t\psi_s$ with the Hamiltonian acting on the wave functions, so we find
\begin{equation}
\partial_tn(\bm{r},t) + \bm{\nabla}\cdot\bm{j}(\bm{r},t) = 0,
\end{equation}
where the second term represents the divergence of the current vector $\bm{j}$, the microscopic representation of which can be derived as
\begin{equation}\label{current density}
j^a(\bm{r},t) = \Bigl\langle\frac{1}{2m_p}\left((\hat{j}^{\ a}_p\psi_s)^{\dagger}\psi_s
+ \psi_s^{\dagger}(\hat{j}^{\ a}_p\psi_s)\right)\Bigr\rangle.
\end{equation}
In the next step, we differentiate the momentum density \eqref{current density} with respect to time and apply the many-particle Pauli-Schrodinger equation to time derivatives of the wave functions as we did above
\begin{align}\label{j}
\partial_tj^{a} + \partial_{b}\Pi^{ab} =& \,\frac{\mu}{m}\bm{s}\cdot\nabla^a\bm{B}
+ F^{a}_{\rm aB} + F^{a}_{\rm as}\nonumber\\ & + \,F^{a}_{\rm asr} + F^a_{\rm asc} + F^a_{\rm ni}.
\end{align}
The momentum flux tensor density $\Pi^{ab}(\bm{r},t)$ can be obtained in the microscopic representation in the process of Eq. \eqref{j}
\begin{equation}\label{current flux}
\Pi^{ab} = \Bigl\langle\!\frac{1}{4m_p}\!\left(\psi^{\dagger}_s(\hat{j}^a_p\hat{j}^b_p\psi_s)
+ (\hat{j}^a_p\psi_s)^{\dagger}(\hat{j}^b\psi_s) + h.c.\right)\!\Bigr\rangle.
\end{equation}
The dynamics of a system of particles in a non-inertial frame of reference and under the action of an external axion field, variable both in time and space, is determined by the force fields on the right-hand side of the equation \eqref{j}. The first force field on the right hand side of this equation represents a well-known Stern-Gerlach force and leads to the separation of the spin-up and spin-down states in the non-uniform magnetic field $\bm{B}$. The second term $\bm{F}_{\rm aB}(\bm{r},t)$ characterizes the spin-axion and spin-axion-rotation coupling effects in an external magnetic field and consists of two terms 
\begin{align}
F^{a}_{\rm aB} = \overbrace{-\,\frac{2\mu}{mc\hbar}\lambda\epsilon^{abc}s^b B^c}^{\substack{
\text{spin-axion coupling} \\ \text{in a magnetic field}}}\quad\quad\quad\quad\quad\quad\quad\nonumber\\
\quad+\overbrace{\,\frac{2\mu}{mc^2\hbar}\epsilon^{abc}\epsilon^{dmn}\lambda^d\cdot
\Lambda^{bm} B^c\omega^n.}^{\substack{\text{space-axion-rotation coupling}\\ \text{in a magnetic field}}}
\end{align}
The force field density $F^a_{\rm as}$ is caused by the coupling of spin with the space non-uniform and time-varying axion field  
\begin{align}
F^a_{\rm as}=\overbrace{-\,\frac{1}{m}s^{b}\cdot \nabla^a\lambda^b}^{\substack{\text{spin-not uniform axion}\\
\text{field coupling}}}-\overbrace{\,\frac{1}{mc}s^{a}\cdot \partial_t\lambda,}^{\substack{\text{spin - time-varying axion} \\ \text{field coupling}}}
\end{align}
and the force field density that is caused by the spin-axion-rotation coupling \begin{align}
F^a_{\rm asr} =&\, \overbrace{-\,\frac{1}{mc}\nabla^a\lambda (\bm{\omega}\cdot\bm{L})}^{
\substack{\text{axion-spin momentum coupling} \\ \text{in rotating frame}}}\overbrace{
+\,\frac{1}{mc^2}\varepsilon^{mbc}\partial_t\lambda^m \omega^b\Lambda^{ac}}^{\substack{
\text{axion field-spin} \\
\text{-rotation coupling}}}\nonumber\\
& \overbrace{+\,\frac{1}{mc^2}\varepsilon^{mbc}\lambda^m
\partial_t\omega^b\Lambda^{ac},}^{\substack{\text{axion field-spin} \\ \text{-not uniform rotation coupling}}}
\end{align}
where the microscopic definition for the mechanical spin moment density $L^a(\bm{r},t)=\bm{L}(\bm{r},t)$ has the form
\begin{equation}
\bm{L} = \biggl\langle\psi_s^{\dagger}\left(\hat{\bm{\sigma}}_p\times\hat{\bm{r}}_p\right)\psi_s\biggr\rangle,
\end{equation}
where $\bm{\lambda}(\bm{r},t)=\frac{c\hbar g_f}{2f_a}\bm{\nabla}a(\bm{r},t)$ is the spatial gradient of a pseudo-scalar axion field, and $\lambda(\bm{r},t) = \frac{\hbar g_f}{2f_a} \partial_ta(\bm{r},t) $.
   
The fifth force field term $F^{a}_{\rm asc}(\bm{r},t)$ arises from the coupling of the spin current density with the pseudo-scalar component of the axion field $\lambda(\bm{r},t)$
\begin{align}\label{asc}
F^{a}_{\rm asc}=\frac{1}{mc}\biggl(\nabla^{a}\lambda\cdot j^{bb}_s
-\nabla_{b}\lambda\cdot j^{ab}_s\biggr).
\end{align}
As we can see, the force field \eqref{asc} has a complex form and is caused by the influence of the inhomogeneous axion field on the spin current density $j_s^{ab}(\bm{r},t)$. It should be noted that we exclude from consideration and don't take into account force fields and torques in Eq. \eqref{spin equation}, \eqref{j} that are proportional to the square of the angular velocity of rotation $\sim \omega^2$ in the regime of slow rotation.

Inertial forces are characterized by the field
\begin{equation}
\bm{F}_{\rm ni} = \bm{R}\times\partial_t\bm{\omega}+2\bm{j}\times\bm{\omega}
- m\bm{\omega}\times(\bm{\omega}\times\bm{R}),
\end{equation}
where $\bm{R} = \langle\psi_s^{\dagger}\hat{\bm{r}}_p\psi_s\rangle$.

\subsection{Macroscopic representation of the momentum flux tensor and spin current density}

In this section, we derive the macroscopic representation of the momentum flux tensor density $\Pi^{ab}$ and spin current tensor density $j^{ab}_s$ in terms of the observable macroscopic variables such as the concentration of particles $n(\bm{r},t)$, the macroscopic average fluid velocity $\bm{\upsilon}(\bm{r},t)$ and spin density $\bm{s}(\bm{r},t)$ (the macroscopic average spin $\bm{S}(\bm{r},t)$). The macroscopic fluid description was developed in detail in Ref. \cite{Trukhanove_2013}. Here we present the key results of the decomposition of physical quantities. To begin with, we recall the representation of the many-particle wave function in exponential form, or the so-called Madelung decomposition of the $N$-particle wave function $\psi_s(R,t) = \rho(R,t)\exp\left(\frac{iS(R,t)}{\hbar}\right)\varphi(\bm{r},R,t)$, where $\rho(R,t)$ is an amplitude, $S(R,t)$ is a phase, and $\varphi$ is a normalized spinor ($\varphi^{\dagger}\varphi=1$). This underlies the subsequent decomposition of the basic variables (such as flow velocity and spin density) into the sum of macroscopic averages and thermal fluctuations.

The momentum current density tensor or the momentum flux tensor $\Pi^{ab}(\bm{r},t)$ may be obtained in macroscopic form of decomposition
\begin{equation}\label{Omega}
    \Pi^{ab} = mn\upsilon^a\upsilon^b + p^{ab} + T^{ab} + \sigma^{ab},
\end{equation}
where the second term is the kinetic pressure tensor $p^{ab}(\bm{r},t)$ in the microscopic representation \cite{Trukhanove_2013, Andreev2013OnAC}
\begin{align}
p^{ab}(\bm{r},t) &= \biggl\langle \rho^2m_pu^a_p u^b_p\biggr\rangle \nonumber\\
&= \int dR\sum_p^N \delta(\bm{r} - \bm{r}_p)\rho^2(R,t)m_pu^a_p u^b_p, 
\end{align}
and the Bohm quantum potential $T^{ab}(\bm{r},t)$, which  for the large system of non-interacting particles can be presented as
\begin{align}
T^{ab}(\bm{r},t) &= -\,\biggl\langle \rho^2(R,t)\partial_p^a\partial_p^b \ln [\rho(R,t)]\biggr\rangle\nonumber\\
&= -\,\frac{\hbar^2}{4m}\partial^a\partial^b n + \frac{\hbar^2}{4m}\frac{1}{n}(\partial^a n) (\partial^b n).
\end{align}
The spin part of the Bohm potential $\sigma^{ab}(\bm{r},t)$, which produces a force by the self-interaction of the spin particles, can be derived as
\begin{align}
\sigma^{ab}(\bm{r},t) &= \frac{\hbar^2}{4m}\biggl\langle \rho^2
\partial_p^a s_p^c \partial_p^b s^c_p\biggr\rangle \nonumber\\
&= \frac{\hbar^2}{4m}n\,\nabla^a (\bm{s}/n)\cdot\nabla^b (\bm{s}/n) + Q^{ab}_{\rm thermal}.
\end{align}
The expression for the momentum current density tensor \eqref{Omega} can be derived using the macroscopic fluid description, which is based on the explicit separation of the movement of particles $\bm{u}_p(\bm{r},R,t) = \bm{\upsilon}_p(R,t) - \bm{\upsilon}(\bm{r},t)$ with thermal velocities $u_p^a(\bm{r},R,t)$ and the collective movement of particles with velocity $\upsilon^a(\bm{r},t)$ in the momentum flux density \eqref{current flux}. The thermal fluctuation of the spin about the macroscopic average $\bm{s}(\bm{r},t)$ can also be decomposed $\bm{w}_p(\bm{r},R,t) = \bm{s}_p(R,t) - \bm{S}(\bm{r},t)$. We also take into account that the average values of the thermal velocity and fluctuations of the spin are equal to zero $\langle\rho^2\upsilon^a_p\rangle = 0$, and $\langle\rho^2w^a_p\rangle = 0$. The current density and spin density can be represented by the velocity of the fluid $\bm{j}=n\bm{\upsilon}$, $\bm{s}=n\bm{S}$.
 
The microscopic definition for the spin current density can be represented in the form
\begin{align}
j^{ab}_s(\bm{r},t) &= \biggl\langle \rho^2s_p^a\upsilon_p^b - \frac{\hbar}{2m_p}\rho^2
\epsilon^{acd}s_p^c\nabla_p^b s_p^d\biggr\rangle \nonumber\\
&= s^a\upsilon^b - \frac{\hbar}{2m}\frac{1}{n}\epsilon^{acd}s^c\nabla^b s^d
+ \Theta^{ab}_{\rm thermal}.\label{spin current 2}
\end{align}
The expressions \eqref{Omega} and \eqref{spin current 2} contain thermal-spin interactions $Q^{ab}_{\rm thermal}$ and $\Theta^{ab}_{\rm thermal}$, which are not discussed in detail in the context of this article.

\section{\label{Discussion} Discussion}

It is useful to compare the development of the formalism described above with the findings in the literature. Here we have considered a system of particles with spins in a non-inertial rotating frame of reference and in the presence of external electromagnetic and the pseudoscalar field of dark matter axions. Earlier, the semi-classical equations of motion for an electron in a rotating frame were obtained in Refs. \cite{PhysRevLett106076601,PhysRevB84104410} on the basis of the Pauli-Schr\"odinger equation. The effects of spin coupling and mechanical rotations were analyzed \cite{PhysRevB84104410}, when spin-dependent inertial forces in the presence of electromagnetic fields were deduced from the general covariant Dirac equation. In our paper, we derive the Pauli-Schr\"odinger equation in a rotating frame by using the Foldy-Wouthuysen transformation. The resulting general Hamiltonian \eqref{total H} accounts not only for the influence of the spin-rotation coupling energy, but also for the spin-axion coupling effects. Based on this Hamiltonian, we construct the non-relativistic many-particle formalism to describe the collective dynamics of a system of quantum particles and introduce the many-particle Pauli-Schr\"odinger equation with the Hamiltonian \eqref{FW}. More specifically, later in the article we focus on the case of neutral particles with spin in a rotating frame and a background axion field and establish the general structure of the spin-dependent force densities in a rotating frame in the presence of an axion field and an external magnetic field. This result is essentially new.

The dynamical equation of momentum balance \eqref{j} was obtained in the framework of the self-consistent formalism of many-particle quantum hydrodynamics. It describes the collective evolution of motion of particles, as compared to the equation of motion in Refs. \cite{PhysRevLett106076601,PhysRevB84104410} that was obtained for one electron in the non-inertial frame. On the other hand, equation \eqref{j} contains inertial force fields, which in the one-particle approximation coincide with the result of Ref. \cite{PhysRevB84104410}. In the absence of rotation ($\bm{\omega} = 0$), the forces on the right hand side of equation \eqref{j} reproduce the previous results for an inertial frame, whereas the force density fields $F^{a}_{\rm aB},$ $F^a_{\rm as},$ $F^a_{\rm asr}$ and $F^{a}_{\rm asc}$ describe new effects. In particular, it is worthwhile to notice that the fluid dynamical equation contains the analogue of the Stern-Gerlach force $\sim s_b\cdot(\bm{\nabla}\lambda^b)$ for a general case of the spatially-nonuniform background axion field. In addition, the force field $\sim \lambda\bm{s}\times\bm{B}$ is proportional to the vector product of spin density and magnetic field  in the time-varying axion field, and it accounts for the analogue of the spin-Hall effect in the axion field. Yet another force field appears only in the rotating frame $\sim \nabla^a\lambda (\bm{\omega}\cdot\bm{L})$, being directed along the gradient of the axion field and it vanishes when the rotation velocity is perpendicular to the density of the angular momentum of the spin $\bm{\omega}\perp\bm{L}$. 

The spin evolution equation, or the Landau-Lifshitz-Bloch equation, is commonly used to analyze interactions of fermions with axion fields. In addition to the torques caused by the external magnetic field and the spin-rotation coupling, the spin evolution equation \eqref{spin equation} describes the effect of  the torque on the spin density $\sim \bm{s}\times\bm{\lambda}$, which arises from the axion field. Of greatest interest are the two new torques due to the spin-axion coupling, derived for the first time in the spin density balance equation. The first torque $\sim \dot{a}\varepsilon^{abc}j^{bc}_s$ is determined by the spin current density and is non-trivial for the time-varying axion field. The second torque  $\sim \dot{a} \Lambda^{ab}$ is proportional to the new spin-orbit tensor (\ref{Lam}) and appears only in a rotating frame of reference.

\section{Conclusion}

In this paper, a general formalism was developed for a system of many particles with spin in the non-inertial reference system, rotating with the angular velocity $\bm{\omega}$, and an external axion field $a(\bm{r},t)$ that can be spatially non-uniform and change arbitrarily in time. The Foldy-Wouthuysen transformation leads to the total Hamiltonian \eqref{total H}, which accounts for the influence of the magnetic field on magnetic moments, or Zeeman energy, the spin-rotation and spin-axion coupling, and the relativistic correction, which characterizes the energy of the spin-orbit coupling in the rotating frame of reference. In order to study in greater detail the influence of inertial effects in an external axion field, we explore the dynamics of fermions for the case of zero electric field. We consider the collective dynamics of a system of many neutral particles with spins, and apply the method of many-particle quantum hydrodynamics to obtain a closed system of balance equations for the macroscopic physical observables. Among other physical quantities, we study the evolution of the spin density \eqref{spin density} and the current density \eqref{current density}, which are defined from microscopic variables as the quantum-mechanical averages of the corresponding operators. The equation for the evolution of the spin density \eqref{spin equation} encompasses on the right-hand side the torque densities that are associated with the spin-rotation coupling and also characterize the influence of the gradient $\bm{\nabla}a(\bm{r},t)$ and the time-varying axion fields $\dot{a}(\bm{r},t)$ on the spin density $\bm{s}(\bm{r},t)$. The equation of momentum balance \eqref{j}, or current density dynamics, is not typical for the study of spin-axion coupling, but the analysis of this equation is necessary for understanding of non-equilibrium dynamic processes in a system of a large number of particles in external axion fields.

%

\end{document}